\documentclass[10pt,aps,prc,floatfix,twocolumn,superscriptaddress,nofootinbib,preprintnumbers]{revtex4-1}
\usepackage{amsmath,amsfonts,amssymb}
\usepackage{graphicx}
\usepackage{dcolumn}
\usepackage{xspace}
\usepackage{isotope}
\usepackage{physics, isotope}
\usepackage{braket}
\usepackage{bm}
\usepackage[normalem]{ulem}
\usepackage[pdfpagelabels, pdfencoding=auto, psdextra]{hyperref}
\hypersetup{%
 pdfsubject=Paper,
 pdfkeywords={nuclear reactions} {Bayesian analysis} {cross sections} {machine learning},
 unicode = true,
 breaklinks = true,
 colorlinks = true,
 linkcolor = blue,
 menucolor = blue,
 citecolor = blue,
 urlcolor = blue
}

\graphicspath{{./figs/}}

\newcommand{\eg}{\textit{e.g.}\xspace}
\newcommand{\ie}{\textit{i.e.}\xspace}
\newcommand{\etal}{\textit{et~al.}\xspace}

\newcommand{\psitrial}{\psi_\mathrm{trial}}

\newcommand{\Utilde}{\widetilde{\vb{U}}}

\newcommand{\fm}{\, \text{fm}}

\newcommand{\fmis}{\, \text{fm}^{-2}}

\newcommand{\MeV}{\, \text{MeV}}

\newcommand{\NNLO}{\ensuremath{{\rm N}{}^2{\rm LO}}\xspace}

\newcommand{\article}{article\xspace}

\renewcommand{\vec}{\bm}

\allowdisplaybreaks

\begin{document}

\title{Toward emulating nuclear reactions using eigenvector continuation}

\author{C.~Drischler}
\email{drischler@frib.msu.edu}
\affiliation{Facility for Rare Isotope Beams, Michigan State University, East Lansing, MI~48824, USA}

\author{M.~Quinonez}
\email{quinonez@nscl.msu.edu}
\affiliation{Facility for Rare Isotope Beams, Michigan State University, East Lansing, MI~48824, USA}
\affiliation{Department of Physics and Astronomy, Michigan State University, East Lansing, MI~48824-1321, USA}

\author{P.~G.~Giuliani}
\email{giulianp@frib.msu.edu}
\affiliation{Facility for Rare Isotope Beams, Michigan State University, East Lansing, MI~48824, USA}
\affiliation{Department of Statistics and Probability, Michigan State University, East Lansing, MI~48824, USA}

\author{A.~E.~Lovell}
\email{lovell@lanl.gov}
\affiliation{Theoretical Division, Los Alamos National Laboratory, Los Alamos, NM~87545, USA}

\author{F.~M.~Nunes}
\email{nunes@frib.msu.edu}
\affiliation{Facility for Rare Isotope Beams, Michigan State University, East Lansing, MI~48824, USA}
\affiliation{Department of Physics and Astronomy, Michigan State University, East Lansing, MI~48824-1321, USA}

\date{\today}
\preprint{LA-UR-21-25457}

\begin{abstract} 

We construct an efficient emulator for two-body scattering observables using the general (complex) Kohn variational principle and trial wave functions derived from eigenvector continuation.  The emulator simultaneously evaluates an array of Kohn variational principles associated with different boundary conditions, which allows for the detection and removal of spurious singularities known as Kohn anomalies. When applied to the $K$-matrix only, our emulator resembles the one constructed by Furnstahl~\emph{et~al.} [Phys.~Lett.~B~\textbf{809}, 135719] although with reduced numerical noise. After a few applications to real potentials, we emulate differential cross sections for $^{40}$Ca$(n,n)$ scattering based on a realistic optical potential and quantify the model uncertainties using Bayesian methods. These calculations serve as a proof of principle for future studies aimed at improving optical models.

\end{abstract}

\maketitle

\section{Introduction} \label{sec:intro}

There are many reasons to study rare isotopes today; \eg, they play a crucial role in obtaining a fundamental understanding of the nucleosynthesis of heavy, neutron-rich nuclei and the dense matter inside neutron stars~\cite{Balantekin:2014opa, Nunes:2020bue, Drischler:2021kxf}.
Due to their short lifetimes, rare isotopes are primarily investigated through reaction experiments conducted at radioactive beam facilities worldwide, including RIKEN, FAIR, GANIL, and soon also FRIB. 
For the analysis and interpretation of these experiments, reliable reaction theory is imperative.
However, apart from reactions on light nuclei, reaction theory is still largely phenomenological and relies on poorly constrained effective interactions to keep calculations tractable~\cite{Johnson_2020}.

Statistical methods such as Bayesian parameter estimation~\cite{Wesolowski:2018lzj} and model comparison~\cite{Phillips:2020dmw} can provide important insights into the issues of effective interactions. They can also help design next-generation reaction experiments (see, \eg, Ref.~\cite{Melendez:2020ikd}). 
But in practice their applications are limited because Monte Carlo sampling of the models' parameter spaces in reaction calculations is usually computationally demanding.
Hence, Bayesian studies of nuclear reactions~\cite{lovell2018,king2019,catacora2019} have only considered the simplest reaction theory, the optical model, which describes, \eg, nucleus-nucleus scattering as two particles interacting via a complex-valued interaction.
Extending these studies to more sophisticated reaction theories (see, \eg, Refs.~\cite{capel2012,diego2014,summers2006,summers2014, alvarez2010,nguyen2012,Lovell2017, Deltuva:2009xg,hlophe2019}) is a challenging yet important task. 

Emulators---computationally inexpensive algorithms capable of approximating exact model calculations with high accuracy---are promising tools in this regard~\cite{sangaline2016,Neufcourt2018,kravvaris2020,Lovell2020}. 
In particular, eigenvector continuation (EC)~\cite{Frame:2017fah,sarkar2021} has been shown to be a powerful method for emulating bound-state properties such as binding energies and charge radii of atomic nuclei~\cite{Konig:2019adq, Ekstrom:2019lss, Wesolowski:2021cni}.
Furnstahl~\etal~\cite{Furnstahl:2020abp} have recently demonstrated that EC also allows for the construction of effective trial wave functions for calculations of two-body scattering observables using the $K$-matrix Kohn variational principle (KVP)~\cite{Kohn:1948col}.
Further, Melendez~\etal~\cite{Melendez:2021lyq} have extended the EC concept to trial $K$- or $T$-matrices in applications of Newton's variational method to two-body scattering, \eg, with a modern chiral interaction.
Remarkably high accuracies and speedups relative to exact scattering calculations were obtained~\cite{Furnstahl:2020abp,Melendez:2021lyq}.
(See Ref.~\cite{bai2021} for EC applied to $R$-matrix theory calculations of fusion observables.)

In this \article, we improve and extend the emulator developed by Furnstahl~\etal~\cite{Furnstahl:2020abp} in several ways. 
Besides the $K$-matrix, we emulate a variety of matrices associated with different scattering boundary conditions simultaneously via the general (complex) KVP~\cite{Lucchese:1989zz}. 
(For pre-EC studies with this method, including nucleon-deuteron scattering, see Refs.~\cite{Kievsky:1997zf,Kievsky99,Lombardi:2004eu}.)
This approach allows us to detect and remove spurious singularities known as Kohn anomalies~\cite{PhysRev.124.1468,nesbet2013variational}, which can render variational calculations of scattering observables ineffective---especially when used for sampling a model's parameter space. 
We also propose a method for solving the emulator equations in Ref.~\cite{Furnstahl:2020abp} with reduced numerical noise.
As a step toward emulating nuclear reactions, we apply our emulator to differential cross sections in \isotope[40]{Ca}$(n,n)$ scattering using a realistic optical potential and quantify the uncertainties in the model parameters using Bayesian methods.

The remainder of this \article is organized as follows. 
In Section~\ref{sec:gKVP} we introduce the formalism of the general KVP with EC trial wave functions. 
We then present several applications of our emulator to realistic potentials (including a chiral potential) in Section~\ref{sec:results}. 
Section~\ref{sec:summary_outlook} concludes the \article with a summary and outlook.
Additional information, \eg, on redundancies among different KVPs with EC trial wave functions, is provided in Appendices~\ref{sec:app:moebius} to~\ref{sec:app:add_results}.
We use natural units in which $\hbar = c = 1$.

\section{Formalism} \label{sec:gKVP}

We consider here local short-range potentials $V(\vb*{\theta})$ in coordinate space that depend on a set of free parameters $\vb*{\theta}$; \eg, the parameters of an optical model or low-energy couplings of a chiral potential. 
Further, $V(\vb*{\theta})$ is assumed to be partial-wave decomposed into an uncoupled channel with angular momentum $\ell$. 
Following Furnstahl~\etal~\cite{Furnstahl:2020abp}, we then use EC to construct an effective trial wave function for our (nonrelativistic) variational calculations of two-body scattering observables:
\begin{equation} \label{eq:nEC}
    \ket{\psi_\mathrm{trial}} = \sum_{i=1}^{N_b} c_{\ell,E}^{(i)} \ket{\psi_{\ell,E}(\vb*{\theta}_i)} \,.
\end{equation}
Here, each of the $N_b$ basis wave functions, \ie, 
\begin{equation} \label{eq:partialwave}
    \innerproduct{\vec{r}}{{\psi_{\ell,E}(\vec{\theta}_i)}} = \frac{\phi_{\ell,E}^{(i)}(r)}{r} Y_{\ell}^{m}(\Omega_r) \,,
\end{equation}
%
is an exact (partial-wave) solution to the Schr{\"o}dinger equation for $V(\vb*{\theta}_i)$ at the center-of-mass energy ${E>0}$, and the coefficients $c_{\ell,E}^{(i)}$ are to be determined. The radial wave functions in Eq.~\eqref{eq:partialwave} are normalized by imposing asymptotic boundary conditions\footnote{The boundary condition~\eqref{eq:asymLimitParam} can be extended to the (long-range) Coulomb potential. See, \eg, Eq.~(S11) in Ref.~\cite{Furnstahl:2020abp}.} of the general form~\cite{Lucchese:1989zz}
\begin{equation} \label{eq:asymLimitParam}
    \phi_{\ell,E}(r) \sim  \bar{\phi}_{\ell,E}^{(0)}(r) + L_{\ell,E} \, \bar{\phi}_{\ell,E}^{(1)}(r) \,, 
\end{equation}
where the two independent free-space solutions are expressed in terms of a nonsingular (complex) matrix $\vb{u}$ that is associated with the generic $L$-matrix in Eq.~\eqref{eq:asymLimitParam}:
\begin{align} 
\begin{pmatrix}
    \bar{\phi}_{\ell,E}^{(0)}(r)\\
    \bar{\phi}_{\ell,E}^{(1)}(r)
\end{pmatrix} &\sim
\mathcal{N}^{-1}
\begin{pmatrix}
    u_{00} & u_{01}\\
    u_{10} & u_{11}\\
\end{pmatrix}
\begin{pmatrix}
    \sin \eta_\ell(r) \\
    \cos \eta_\ell(r)
\end{pmatrix} \label{eq:matrixparam} \,,
\end{align}
with $\eta_\ell(r) = pr - \frac{\pi}{2}\ell$ and $p = \sqrt{2\mu E}$, and an arbitrary normalization constant $\mathcal{N} \neq 0$. For instance, the familiar $K$-, $S$-, and $T$-matrix respectively correspond to\footnote{The matrices are determined only up to scalar multiples. Lucchese~\cite{Lucchese:1989zz} uses a different convention for the $S$-matrix ($u\to-u$) and $T$-matrix parametrization ($T\to-\pi T$).}
\begin{equation} \label{eq:uMatrices}
    \vb{u} = \begin{pmatrix} 1 & 0\\ 0 & 1 \end{pmatrix}\qc
    \vb{u} = \begin{pmatrix} -i & 1\\ -i & -1 \end{pmatrix}\,, \qq{and}
    \vb{u} = \begin{pmatrix} 1 & 0\\ i & 1 \end{pmatrix}.
\end{equation}
But any other nonsingular parametrization $(L,\vb{u})$ of the asymptotic limit~\eqref{eq:asymLimitParam} is equally valid.\footnote{For example, by swapping the rows in $\vb{u}$ associated with $L$ [\eg, given by Eq.~\eqref{eq:uMatrices}] one obtains the matrix $\vb{u}'$ parametrizing $L^{-1}$.} The corresponding $K$-matrix can be obtained using the M{\"o}bius transformation (see also Appendix~\ref{sec:app:moebius})
\begin{equation} \label{eq:trafo_K_L}
    K_{\ell,E}\left(L_{\ell,E}\right) = \frac{u_{01} + u_{11} L_{\ell,E}}{u_{00} + u_{10} L_{\ell,E}} \,,
\end{equation}
which is related to the phase shift via $K_{\ell,E} = \tan \delta_{\ell,E}$.

In Appendix~\ref{sec:app:training} we give the technical details for solving the (radial) Schr{\"o}dinger equation numerically to determine the basis wave functions in Eq.~\eqref{eq:nEC}, including a generalization of Eq.~\eqref{eq:trafo_K_L} to transform from a given $L$-matrix to any other $L'$-matrix---not just the $K$-matrix. 

We determine the coefficients $c_{\ell,E}^{(i)}$ in Eq.~\eqref{eq:nEC} using the general (complex) KVP. Given $(L,\vb{u})$ and a trial wave function $\innerproduct{\vec{r}}{{\psitrial}}$ subject to the boundary condition~\eqref{eq:asymLimitParam}, the general KVP provides a stationary approximation to $\tilde{L}_{\ell,E} = L_{\ell,E}/\mathcal{N}$ using the functional~\cite{Lucchese:1989zz}\footnote{Lucchese~\cite{Lucchese:1989zz} considers electron-nucleus scattering in atomic units, wherein the electron mass $m_e = 1$ and thus $\mu\approx 1$. For optical potentials the bra-states are to be complex conjugated (as indicated by the asterisk).}
\begin{equation}\label{eq:betatrial}
    \beta_{\vb{u}}\left[ \ket{\psitrial}\right] = \tilde{L}_{\ell,E} - \frac{\mathcal{N}}{p} \frac{2\mu}{\det \vb{u}}  \Braket{\psitrial^{*}|H(\vb*{\theta})-E|\psitrial}\,,
\end{equation}
with the reduced mass $\mu \approx A_p A_t/(A_p + A_t) \, m_n$, mass number of the projectile $A_p=1$ (here, a neutron with mass $m_n$) and target $A_t$, respectively, as well as the Hamiltonian $H(\vb*{\theta}) = -\grad^2/(2\mu) + V(\vb*{\theta})$ in coordinate space. A derivation similar to the one in Refs.~\cite{taylor2006scattering,Furnstahl:2020abp} for the $K$-matrix shows that the functional~\eqref{eq:betatrial} is indeed stationary about exact solutions to the Schr{\"o}dinger equation, 
\ie, $\beta_{\vb{u}}[ \ket{\psi_{\ell,E}} + \ket{\delta \psi_{\ell,E}} ]= \tilde{L}_{\ell,E} + (\delta \tilde{L}_{\ell,E})^2$,
although it does not provide an upper or lower bound in general. 

We impose the normalization constraint $\sum_{i} c_{\ell,E}^{(i)} = 1$ on the EC trial wave function~\eqref{eq:nEC} to fulfill the boundary condition~\eqref{eq:asymLimitParam} required by the general KVP.
Constrained optimization of the functional~\eqref{eq:betatrial} using a Lagrange multiplier $\lambda$ then leads to the system of $(N_b+1)$ linear equations
\begin{equation}\label{eq:kvpLSsolver}
    \begin{pmatrix}
\Delta \Utilde_{\ell,E}^{(\vb{u})} & \vb{1}\\
\vb{1}^\textsf{T} & 0
\end{pmatrix}
\begin{pmatrix}
\vb{c}_{\ell,E}\\
\lambda
\end{pmatrix}
=
\begin{pmatrix}
\vb{\widetilde{L}}_{\ell,E}\\
1
\end{pmatrix}\,,
\end{equation}
and eventually to the desired stationary solution~\cite{Furnstahl:2020abp} 
\begin{subequations} \label{eq:kvpSolAna}
\begin{align}
c_{\ell,E}^{(i)} &= \sum_{j=1}^{N_b} \left(\Delta\Utilde^{(\vb{u})}_{\ell,E}\right)^{-1}_{ij} \left( \tilde{L}_{\ell,E}^{(j)}  - \lambda\right), \label{eq:kvpSolci} \\
\lambda &= \frac{-1 + \sum \nolimits_{i,j=1}^{N_b} \left(\Delta\Utilde_{\ell,E}^{(\vb{u})}\right)^{-1}_{ij} \tilde{L}_\ell^{(j)}}
         {\sum \nolimits_{i,j=1}^{N_b} \left(\Delta\Utilde_{\ell,E}^{(\vb{u})}\right)^{-1}_{ij}}\,. \label{eq:kvpSolLambda}
\end{align}
\end{subequations}
In Eq.~\eqref{eq:kvpLSsolver}, $\vb{1}$ ($\vb{1}^\textsf{T}$) is an all-ones column vector (row vector), and $\vb{\widetilde{L}}_{\ell,E}$ and $\vb{c}_{\ell,E}$ are vectors respectively containing the $L$-matrices of the basis wave functions and the unknown coefficients~$c_{\ell,E}^{(i)}$. 
Further, we have defined the $N_b\times N_b$ kernel matrix
\begin{subequations} \label{eq:deltaUTilde}
\begin{align}
\left( \Delta \Utilde_{\ell,E}^{(\vb{u})}\right)_{ij} &= \frac{\mathcal{N}}{p} \frac{2\mu}{\det \vb{u}} \left[ 2 A_{ij} - B_{ij} \right] \,, \label{eq:def_deltaUTilde}\\ 
A_{ij} &= \Braket{\psi_{\ell,E}^*(\vb*{\theta}_i)|V(\vb*{\theta}) |\psi_{\ell,E}(\vb*{\theta}_j)} \,, \label{eq:Amat} \\ 
B_{ij}  &= \Braket{\psi_{\ell,E}^*(\vb*{\theta}_i)| V(\vb*{\theta}_i) + V(\vb*{\theta}_j) |\psi_{\ell,E}(\vb*{\theta}_j)} \,, \label{eq:Bmat}
\end{align}
\end{subequations}
which can be efficiently evaluated for a variety of different $(L, \vb{u})$ at once, as discussed in Appendix~\ref{sec:app:kernel}. Hence, the stationary approximation to $L_{\ell,E}$ reads
\begin{equation}\label{eq:Lapprox}
       [L_{\ell,E}]_{\text{KVP}} =
    \sum \limits_{i=1}^{N_b} c_{\ell,E}^{(i)} L_{\ell,E}^{(i)}
    - \frac{\mathcal{N}}{2} \sum \limits_{i,j=1}^{N_b} c_{\ell,E}^{(i)} \left(\Delta\Utilde_{\ell,E}^{(\vb{u})} \right)_{ij} c_{\ell,E}^{(j)}\,.
\end{equation}
%

Equations~\eqref{eq:kvpLSsolver} to~\eqref{eq:Lapprox} are the main expressions of our emulator. They reduce to the ones derived in Ref.~\cite{Furnstahl:2020abp} for the $K$-matrix KVP (\ie, $L=K$) with $\mathcal{N}=p$. Hence, the discussion of the computational complexity in Ref.~\cite{Furnstahl:2020abp} also applies to our emulator for each $(L, \vb{u})$.

The EC trial wave function~\eqref{eq:nEC} renders the kernel matrix $\Delta \Utilde_{\ell,E}^{(\vb{u})}$ increasingly ill-conditioned as the number of basis wave functions $N_b$ increases. To control the numerical noise due to the explicit matrix inversion in the algebraic solution~\eqref{eq:kvpSolAna}, Furnstahl~\etal~\cite{Furnstahl:2020abp} added a small regularization parameter to the diagonal elements of $\Delta \Utilde_{\ell,E}^{(\vb{u})}$. 
Although this simple approach typically works well in practice, we find that solving instead the system of equations~\eqref{eq:kvpLSsolver} numerically using a least-squares solver [to circumvent explicit matrix inversion] is less sensitive to numerical noise---especially at low energies.
The least-squares solver uses a different regularization method, where singular values less than a given cutoff ratio times the largest singular value are considered zero. In the cases we studied, this cutoff ratio could be as small as the machine epsilon to avoid potential fine-tuning.

In addition to these numerical instabilities, the general KVP is prone to spurious singularities known as Kohn (or Schwartz) anomalies~\cite{PhysRev.124.1468,nesbet1980variational}, which occur at energies where the functional~\eqref{eq:betatrial} does not provide a (unique) stationary approximation~\eqref{eq:Lapprox}. 
(See Section~\ref{sec:real_potentials} for several illustrations.)
For the realistic potentials studied here, we find that neither real KVPs, such as the one for the $K$-matrix, nor complex KVPs, such as the one for the $S$-matrix~\cite{doi:10.1063/1.454462,ADHIKARI1992415}, can guarantee anomaly-free results~\cite{Lucchese:1989zz}.

We therefore emulate a wide range of matrices associated with different scattering boundary conditions simultaneously using the general KVP and assess their consistency. 
As pointed out in Appendix~\ref{sec:app:redundancies}, however, not all KVPs (with EC trial wave functions) provide independent stationary approximations---we derive a simple condition to identify those.
Results that do not pass the consistency checks, \eg, $SS^{-1} = 1$~\cite{Lucchese:1989zz,Viviani:2001sy}, are disregarded by our anomaly detection algorithm and the remaining ones averaged over in an attempt to obtain anomaly-free results.
If none of the KVPs evaluated are consistent, our algorithm iteratively adapts the size of the training set, which usually shifts the Kohn anomalies in each iteration.
We refer to this approach as the ``mixed approach.''
More details on detecting and removing Kohn anomalies are presented in Appendix~\ref{sec:app:diagnostic_tools}.


\section{Results and Discussion}\label{sec:results}

\subsection{Realistic Real Potentials} \label{sec:real_potentials}

We apply our emulator first to three real potentials as test cases.
Specifically, we consider nucleon-nucleon (NN) scattering in the ${^1}\text{S}_0$ channel\footnote{The spectroscopic notation $^{1}\text{S}_{0}$ indicates that the angular momentum $\ell=0$ (``S'') and the total spin $S=0$ of the two nucleons couple to the total angular momentum $J=0$.} based on the Minnesota potential~\cite{THOMPSON197753},
\begin{equation} \label{eq:minnesota}
    V(r) = V_{0R} \, e^{-\kappa_R r^2} + V_{0s} \, e^{-\kappa_s r^2} \,,
\end{equation}
and the local chiral potential at next-to-next-to-leading order (\NNLO) developed by Gezerlis~\etal~\cite{Gezerlis:2014zia} with regulator cutoff $R_0 = 1.0\fm$ and spectral-function cutoff $\tilde{\Lambda} = 1000 \MeV$. 
The Minnesota potential allows for direct comparisons with the emulators constructed in Refs.~\cite{Furnstahl:2020abp,Melendez:2021lyq}, and the chiral potential is commonly used in quantum Monte Carlo calculations of atomic nuclei and nuclear matter (see, \eg, Refs.~\cite{Lynn:2019rdt, Piarulli:2019cqu} for recent reviews). The latter potential depends on 8 parameters (\ie, NN low-energy couplings) in the ${^1}\text{S}_0$ channel.\footnote{Only two independent (spectroscopic) low-energy couplings contribute to the ${^1}\text{S}_0$ channel, which are given by linear combinations of the couplings mentioned in the text. For details see, \eg, Appendix~A in Ref.~\cite{Gezerlis:2014zia}.} 
Both were constructed to reproduce ${^1}\text{S}_0$ scattering phase shifts. 

We also consider the scattering states of  $n+$\isotope[10]{Be} based on the real Woods--Saxon potential with the spin-orbit (LS) term added, \ie,
\begin{equation} \label{eq:woods_saxon}
    V(r) = -V_0 \, f_\text{WS}(r; R, a) + \vec{\ell}\cdot \vec{s}\, \frac{V_\text{LS}}{r} \dv{r} f_\text{WS}(r; R, a)
\end{equation}
with the function
\begin{equation} \label{eq:fWS}
    f_\text{WS}(r; R, a) = \left[1+\exp \left( \frac{r-R}{a} \right)\right]^{-1}\,,
\end{equation}
which was fit in Ref.~\cite{Capel2004} to low-lying states in \isotope[11]{Be}, including the $\text{d}_{5/2}$ resonance.\footnote{The spectroscopic notation $\text{d}_{5/2}$ indicates that the angular momentum $\ell=2$ (``d'') is coupled to a total angular momentum of the valence particle $j=5/2$.}
Equation~\eqref{eq:woods_saxon} is commonly used to describe the interaction of the valence nucleon(s) with the core nucleus in halo nuclei, such as 
$^{11}$Be($n+^{10}$Be) 
in reaction models~\cite{Capel2004,summers2006,capel2012}, or $^{16}$Be($n+n+^{14}$Be) in decay studies~\cite{Lovell2017,Casal2019}. 
We consider here the $\text{d}_{5/2}$ channel (rather than $\text{s}_{1/2}$) because the breakup calculations in Ref.~\cite{Capel2004} identified this channel as the dominant one for this scattering process (see Figure~1 in Ref.~\cite{Capel2004}).

For the Minnesota potential~\eqref{eq:minnesota}, we follow Furnstahl~\etal~\cite{Furnstahl:2020abp} and train our emulator on the set of points $(V_{0R}, V_{0s}) = \{ (0., -291.85)$, $(100., 8.15)$, $(300., -191.85)$, $(300., 8.15) \}$ in units of $\MeV$, while the other (nonlinear) parameters are fixed at their best fit values, \ie, $\kappa_R = 1.487\fmis$ and $\kappa_s = 0.465\fmis$~\cite{THOMPSON197753}. For the other two potentials, we randomly select the training points within a $\pm 20\%$ interval (in the appropriate units) of the parameters' best fit values, as given in Table~I of Ref.~\cite{Gezerlis:2014zia} for the chiral potential ($N_b = 4$) and Table~I of Ref.~\cite{Capel2004} for the Woods--Saxon potential~\eqref{eq:woods_saxon} with fixed $V_\text{LS} = 21 \MeV \fmis$ ($N_b = 6$). In all cases, we emulate the scattering phase shifts at the best fit values. 

\begin{figure*}[htb]
    \includegraphics{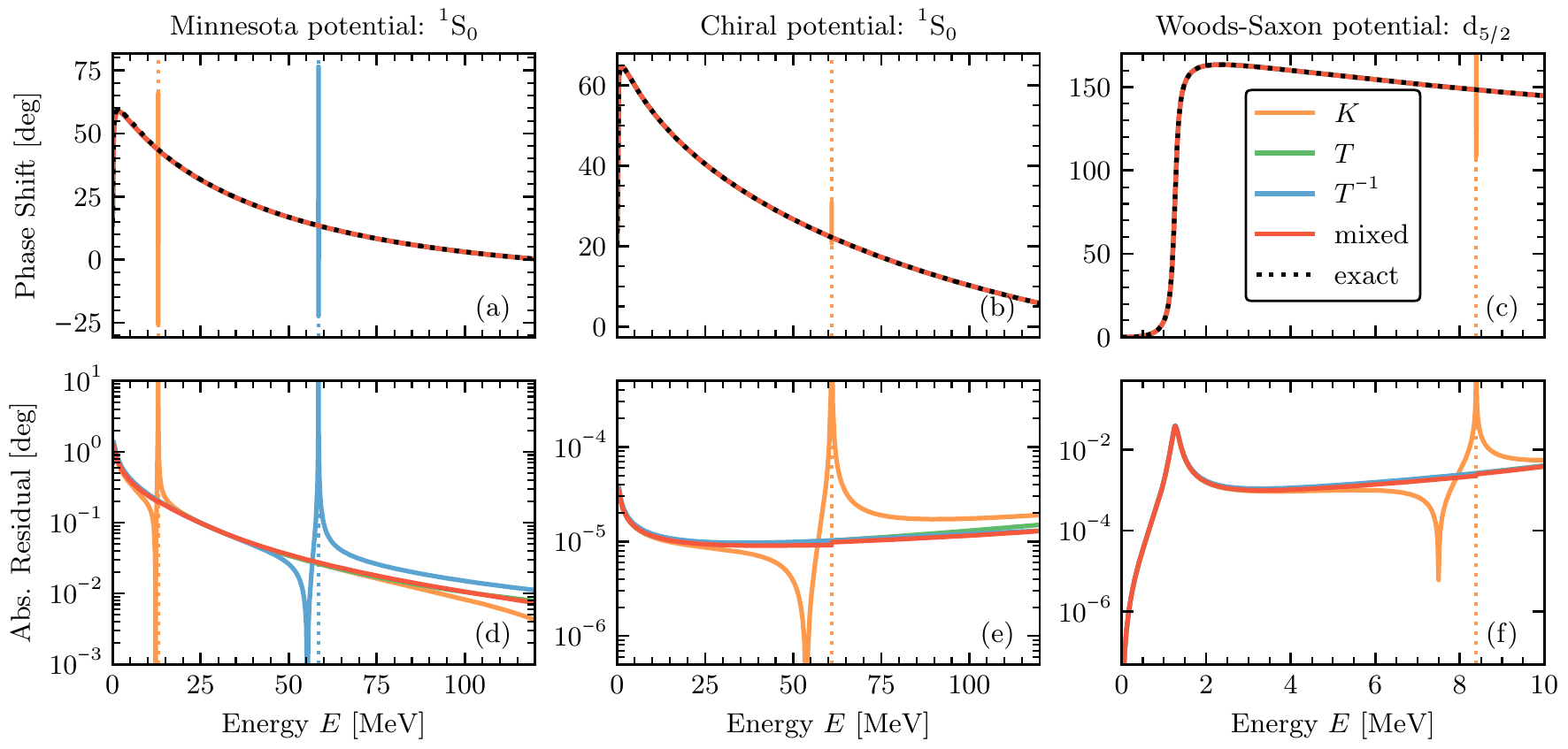}
    \caption{
    Phase shifts (a--c) and the associated absolute residuals (d--f) with respect to the exact solution for the different KVPs (see legend) as a function of the center-of-mass energy: Minnesota potential~\eqref{eq:minnesota} (left column), local chiral potential~\cite{Gezerlis:2014zia} at \NNLO (center column), and Woods--Saxon potential with spin-orbit term~\eqref{eq:woods_saxon} (right column).
    Both, the Minnesota and chiral potential, are used for NN scattering, whereas the Woods--Saxon potential is used for $n+$\isotope[10]{Be} scattering.
    The dotted vertical lines highlight the (approximate) locations of the detected Kohn anomalies. 
    Notice that the algorithm proposed here (red lines) is capable of removing these anomalies.
    See the main text for more details. 
    }
    \label{fig:phaseshifts}
\end{figure*}

Figure~\ref{fig:phaseshifts} shows the emulated phase shifts (a--c) and their absolute residual (d--f) relative to the exact scattering solution as a function of the center-of-mass energy. 
From left to right, the columns correspond to the results obtained for the Minnesota, chiral, and Woods--Saxon potential, respectively. Each panel depicts the emulated results based on the KVPs for the \mbox{$K$-}, \mbox{$T$-}, and $T^{-1}$-matrix, as well as our mixed approach as solid lines. The KVPs for the other canonical matrices (\ie, $K^{-1}$, $S$, and $S^{-1}$) do not provide complementary stationary solutions (as discussed in Appendix~\ref{sec:app:redundancies}) and therefore are not shown. 

Overall, our emulator reproduces well the exact phase shifts. The absolute residuals typically are $\lesssim 0.01^\circ$, except for the Minnesota potential at the low energies where the phase shift is large. As expected, the $K$-matrix KVP (orange lines) reproduces the phase shifts obtained by Furnstahl~\etal~\cite{Furnstahl2021gitrepo} for the Minnesota potential, including the noticeable Kohn anomaly at $E \approx 13 \MeV$. The $T^{-1}$-matrix KVP is anomalous at $E\approx 59 \MeV$. 
In the energy range shown, we also find such an anomaly for the chiral interaction at $E \approx 61 \MeV$, and for the Woods--Saxon potential at $E \approx 8 \MeV$. Additional Kohn anomalies, however, may be present and only noticeable when using extremely fine energy grids~\cite{Furnstahl:2020abp}. Figure~\ref{fig:phaseshifts} emphasizes the need for efficient anomaly removal algorithms beyond proof-of-principle calculations, where the exact scattering solution as a reference is not available.

Such an algorithm is implemented in our emulator (see Section~\ref{sec:gKVP}). Depicted by the red lines in Fig.~\ref{fig:phaseshifts}, the mixed approach is capable of detecting and removing Kohn anomalies by assessing the consistency of the results obtained from a set of different KVPs and (if necessary) adaptively removing basis wave functions from the training set used for emulation. In this specific case, we have simultaneously emulated the complementary matrices $L=(K,T,T^{-1})$, as shown in the figure, as well as the three additional matrices specified in Appendix~\ref{sec:app:diagnostic_tools}. No changes in the training set were necessary to mitigate these Kohn anomalies.

\subsection{Realistic Optical Potential} \label{sec:optical}

We also apply our emulator to a realistic optical potential for \isotope[40]{Ca}$(n,n)$ scattering at $E = 20 \MeV$ in the center-of-mass frame.
Parametrizations of optical potentials (see, \eg, Ref.~\cite{Koning2003}) typically contain real and imaginary terms of the Woods--Saxon form:
\begin{equation}\label{eq:op}
    \begin{split}
        V(r) &= -V_v \, f_\text{WS}(r; R_v, a_v) - i W_v \, f_\text{WS}(r; R_w, a_w) \,,\\
		&\quad - i 4 a_d W_d \, \dv{r} f_\text{WS}(r; R_d, a_d)\,.
    \end{split}
\end{equation}
%
We do not consider the spin-orbit term in Eq.~\eqref{eq:op} and assume in the following 
$R_w=R_v$ and $a_w=a_v$, as in Ref.~\cite{Koning2003}. To train the emulator, we randomly select $N_b$ points for the remaining seven parameters, \ie, $\vb*{\theta} = \{V_v, R_v, a_v, W_v,  W_d, R_d, a_d\}$, within a $\pm 20$\% interval (in the appropriate units) centered around the Koning--Delaroche (KD) parameterization~\cite{Koning2003} at $E = 20 \MeV$. This approach allows us to probe a realistic region of the parameter space.

\begin{figure}[tb]
    \begin{centering}
        \includegraphics[width=\linewidth]{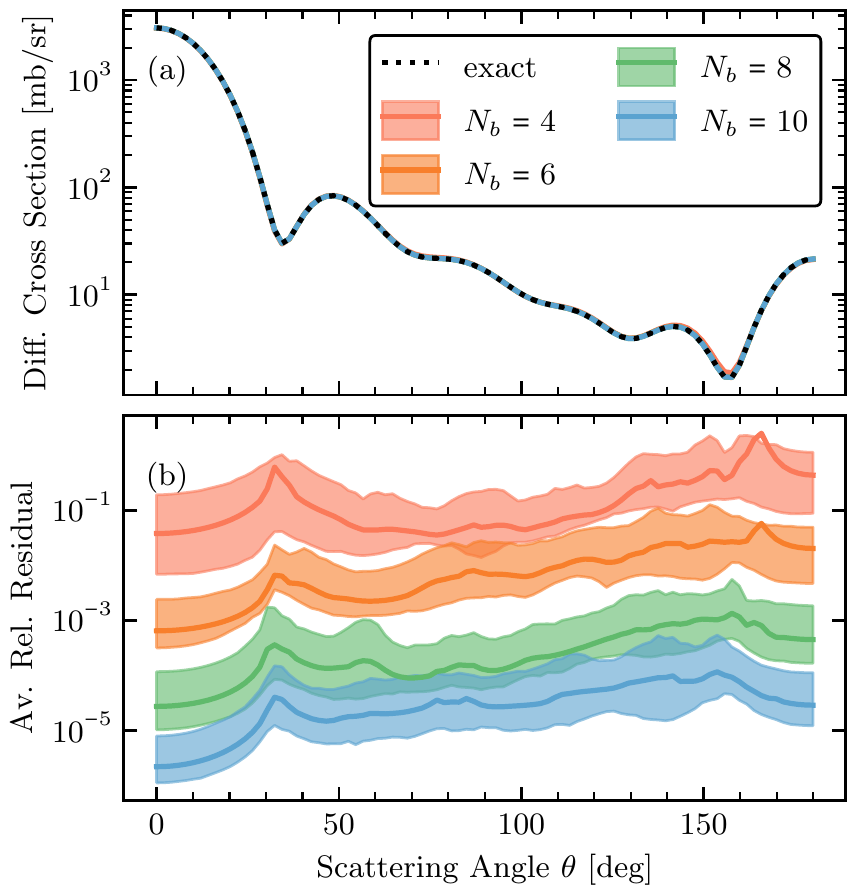}
    \end{centering}
    \caption{Differential cross section~(a) and average relative residual~(b) for $^{40}$Ca$(n,n)$ scattering at $20 \MeV$ using the mixed approach with $500$ random sampling points and four different basis sizes (see legend) as a function of the scattering angle $\theta$. The solid lines depict the average results for the mixed approach using $N_b=4$ (red), $N_b =6$ (orange), $N_b =8$ (green), and $N_b =10$ (blue) basis points. The shaded bands span the range between the 50\% limit (\ie, median) and (upper) 95$\%$ limit of the residuals. The black-dashed line represents the mean value of the exact scattering solutions. For more details see the main text.
    }
    \label{fig:dxs}
\end{figure}

Figure~\ref{fig:dxs} shows~(a) the emulated differential cross sections (mixed approach) and~(b) their corresponding average relative residuals as a function of the scattering angle~$\theta$---which is not to be confused with the parameter set of the interaction, $V(\vb*{\theta})$. 
The exact scattering solutions serve as the reference for the residuals and their mean value is depicted by the black-dotted line in panel~(a). 
We emulate the differential cross section at 500 randomly selected points in the parameter space similar to the training phase, and determine the bands shown in panel~(b) as the range spanned by the 50\% limit (\ie, median) and (upper) 95\% limit of the residuals.
The solid lines in both panels correspond to the average results for the emulators with $N_b=4$ (red lines), $N_b=6$ (orange lines), $N_b=8$ (green lines), and $N_b=10$ (blue lines), respectively. 
We include partial-wave channels with angular momentum $\ell \leqslant 10$ in the calculations.

As shown in Fig.~\ref{fig:dxs}, the accuracy of the emulator roughly improves by an order of magnitude when increasing the size of the training set by increments of two, from $N_b=4$ to $10$.
But the accuracy can also vary by more than an order of magnitude within the 500 sampled points.
Furthermore, increasing the scattering angle tends to decrease the accuracy, which is lowest at the backward angles where the differential cross section is smallest.
Nevertheless, for $N_b\geqslant 6$, the emulator residual does not exceed the experimental uncertainty, typically of the order of $\approx 10$\% (see, \eg, Refs.~\cite{mcdonald1964,vanhoy2018}).
\subsection{Uncertainty Quantification for Optical Models}
\label{sec:UQ}



In this section we explore Bayesian parameter estimation and uncertainty quantification of an optical model using our emulator---as a step toward systematic studies in the future.
For the proof-of-principle calculation we consider again \isotope[40]{Ca}$(n,n)$ scattering at $20 \MeV$\footnote{Additional results for \isotope[40]{Ca}$(n,n)$ scattering at $5 \MeV$ are provided in Appendix~\ref{sec:app:add_results}.} and use the mixed approach with $N_b=8$ training points.
The real and imaginary volume depths and radii of the optical potential (\ie, $V_v$, $R_v$, $W_d$, and $R_d$) are constrained based on mock data generated from the KD potential~\cite{Koning2003} (see Ref.~\cite{Lovell_2020b} for more details), whereas the other optical model parameters are fixed at the original KD values.
Each parameter's prior is taken to be a normal distribution with mean set to the KD potential value  and width of $50\%$ of the mean, similar to previous studies~\cite{lovell2018,king2019,catacora2019}, and the likelihood is the standard exponentiated $\chi^2$.  
The uncertainty quantification is performed through Markov Chain Monte Carlo (MCMC) sampling with 20,000 accepted parameters sets from a single Markov chain.  
We also obtain 95\% confidence intervals for the differential cross sections, defined as the smallest interval over which the posterior distribution integrates to $0.95$.

\begin{figure*}[htb]
\begin{centering}
\includegraphics[width=0.7\linewidth]{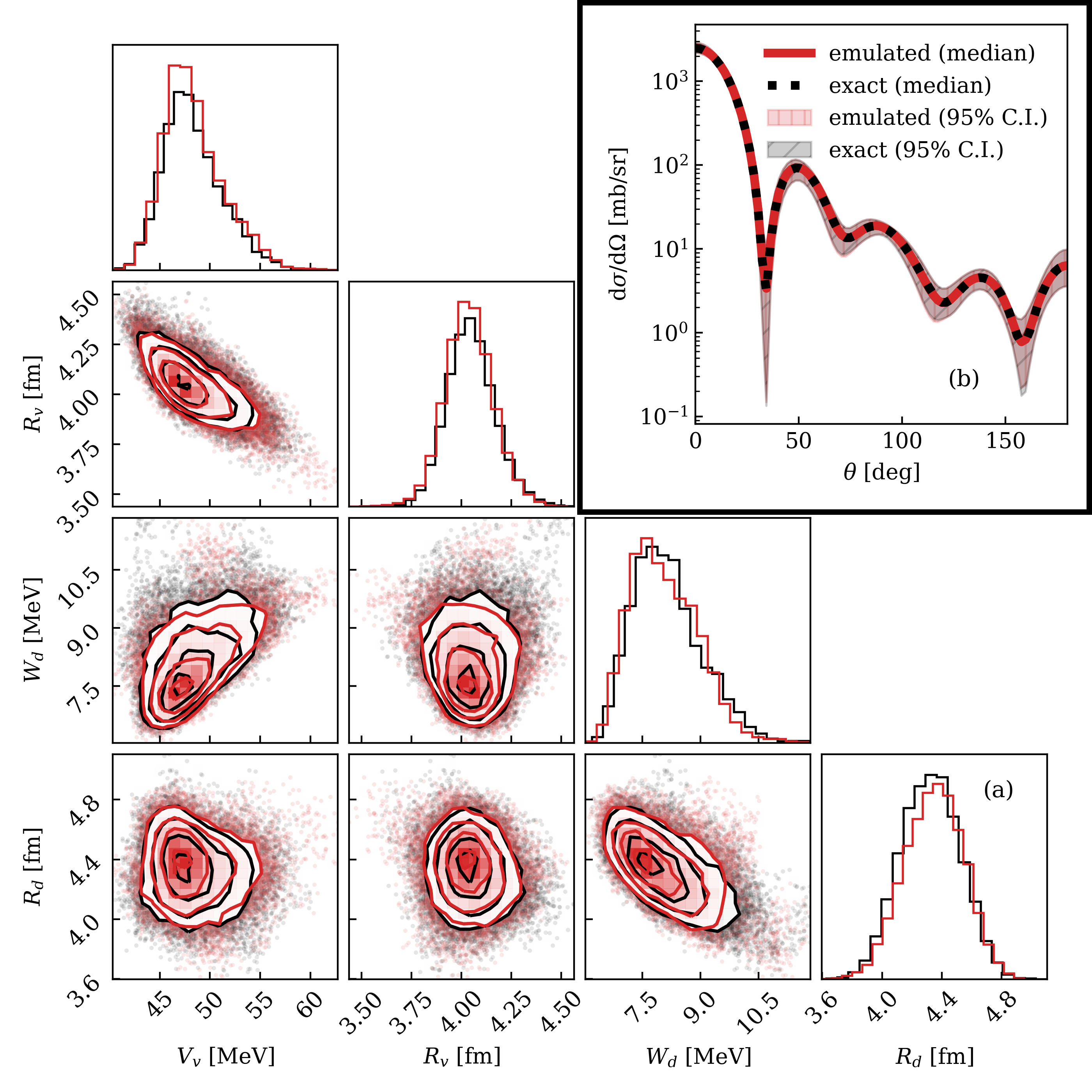}
\end{centering}
\caption{Comparison of the MCMC results obtained for the mixed approach using $N_b=8$ (red) and the exact solution (black) for \isotope[40]{Ca}$(n,n)$ scattering at $20 \MeV$: (a)~posterior distributions for the parameters $V_v, R_v, W_d$, and $ R_d$ along the diagonal, and correlations between each pair of parameters in the off-diagonal; (b)~the median value of the differential cross section as a function of scattering angle (lines) and the corresponding 95\% confidence interval (C.I., shaded area). These MCMC calculations correspond to 20,000 accepted parameters sets.
}
\label{fig:mcmc}
\end{figure*}

Figure~\ref{fig:mcmc} shows the results of the parameter estimation based on the mixed approach (red lines) and the exact scattering solution (black lines). 
Panel~(a) gives the posterior distributions for the four varied parameters along the diagonal, with contour plots displaying the correlations between each pair of parameters in the off-diagonal panels (also known as corner plot).  
Panel~(b) compares the resulting 95\% confidence intervals for emulated vs. exact differential cross sections. 
Apart from statistical fluctuations, the emulator reproduces well the exact calculations of the parameter posterior distributions, correlations, and the confidence intervals for the angular distributions.
Remarkably, our mixed approach obtained anomaly-free results without adapting the training set in all of our MCMC runs.

The mean values of the posterior distributions match the KD parameters, as expected, and the uncertainties of the parameters and the differential cross sections are similar to what has been obtained in previous studies~\cite{lovell2018,Lovell_2020b}. Note that the same reaction has been studied in Ref.~\cite{Lovell_2020b} at slightly lower energy but with a larger set of parameters allowed to vary in the MCMC sampling. 



\section{Summary and Outlook} \label{sec:summary_outlook}

Motivated by the recent Letter by Furnstahl~\etal~\cite{Furnstahl:2020abp}, we constructed an efficient emulator for two-body scattering observables using the general KVP~\cite{Lucchese:1989zz} and trial wave functions derived from EC.
Our emulator does not only consider the $K$-matrix KVP (as in Ref.~\cite{Furnstahl:2020abp}), but rather simultaneously evaluates an array of KVPs associated with different (complex) scattering boundary conditions. 
This approach allows us to systematically detect and remove spurious singularities known as Kohn anomalies, which can render applications of the KVP and other variational principles ineffective; especially when used for Monte Carlo sampling of a model's parameter space.
If only the $K$-matrix KVP is evaluated, our emulator resembles the one constructed in Ref.~\cite{Furnstahl:2020abp} although with reduced numerical noise.

We investigated the EC-driven general KVP in detail and derived analytic expressions to transform the emulator equations of different KVPs efficiently into one another.
In particular, we showed that the stationary solutions of two KVPs are identical if a simple condition is fulfilled [see the discussion of Eq.~\eqref{eq:id_LLp_kvp}]. 

We demonstrated the efficacy of the proposed algorithm for removing Kohn anomalies by emulating scattering phase shifts obtained from the Minnesota, a local chiral, and the real Woods--Saxon potential. 
For each potential, we found anomalies in at least one of the applied KVPs, which the algorithm reliably detected and removed---without adapting the size of the training set. 
This emphasizes that Kohn anomalies need to be dealt with in practice, even in proof-of-principle calculations, but doing so does not require the exact scattering solution.
The basic concept of the algorithm is general and might also be applicable to other variational methods~\cite{Melendez:2021lyq}. 
Furthermore, we showed that, although the emulator's rate of convergence can be sensitive to the details of the interaction and the size of the training set, the high accuracies obtained with our KVP-based emulator are well-suited for scattering calculations. 

After these test applications to real potentials, we studied the EC convergence for emulating differential cross sections in \isotope[40]{Ca}$(n,n)$ scattering at $20 \MeV$ using the realistic KD optical potential. 
A training set with $N_b = 6 - 10$ wave functions, typically, was enough to obtain high-accuracy results for this observable. Next, we performed Bayesian parameter estimation for the optical model by optimizing the emulated differential cross section to reproduce mock data calculated from the KD potential.  
The sampled distribution functions for the model parameters and the differential cross section obtained with the emulator were in excellent agreement with those calculated from the exact scattering solution. 

Important future avenues include the extension of our emulator to scattering in coupled partial-wave and reaction channels, with coordinate and momentum space interactions, as well as the inclusion of the (long-range) Coulomb interaction~\cite{Furnstahl:2020abp,Melendez:2021lyq,taylor2006scattering}.
Technically more challenging will be the extension to emulating three- and higher-body scattering observables, where the computational efficiency of emulators is vital for rigorous uncertainty quantification.
Recent developments in this direction~\cite{Witala:2021xqm,Zhang:2021xx}, however, are promising and will benefit from the insights into the EC-driven general KVP provided here. 
As the number of efficient emulators for scattering observables increases~\cite{Miller:2021pcu,Melendez:2021lyq,Furnstahl:2020abp}, it will be important to benchmark the different emulators quantitatively, \eg, in terms of accuracy, computational speedup, and susceptibility to anomalous behavior. 
These advances set the stage for constructing next-generation optical models using emulators for scattering observables in the FRIB era.

\begin{acknowledgments}

We thank R.~J.~Furnstahl, A.~J.~Garcia, A.~Kievsky, D.~Lee, P.~J.~Millican, T.~R.~Whitehead, and X.~Zhang for fruitful discussions.
We are also grateful to the organizers of the (virtual) INT program ``Nuclear Forces for Precision Nuclear Physics'' (INT--21--1b) for creating a stimulating environment to discuss eigenvector continuation and variational principles.
This material is based upon work supported by the U.S. Department of Energy, Office of Science, Office of Nuclear Physics, under the FRIB Theory Alliance award DE-SC0013617 and award DE-SC0021422. 
This work was carried out under the auspices of the National Nuclear Security Administration of the U.S. Department of Energy at Los Alamos National Laboratory under Contract No. 89233218CNA000001. 
This work was supported by the National Science Foundation under Grant PHY-1811815, and the National Science Foundation CSSI program under Award No. OAC-2004601 (BAND Collaboration~\cite{BAND_Framework}).

\end{acknowledgments}

\appendix
\section{M{\"o}bius transformation}
\label{sec:app:moebius}

The M{\"o}bius (or linear fractional) transformation refers to the function (for more details, see, \eg, Ref.~\cite{2020Fractional})
\begin{equation}\label{eq:moebius_trafo}
    \mathcal{L}_{\vb{a}}(z) = \frac{a_{01} + a_{11} z}{a_{00} + a_{10} z}\,,
\end{equation}
generated by the nonsingular $2\times 2$ matrix $\vb{a}$. We have chosen the order in which the coefficients $a_{ij}$ appear such that Eq.~\eqref{eq:trafo_K_L} reads $K(L) \equiv \mathcal{L}_{\vb{u}}(L)$. If $\vb{a}$ was singular (\ie, $\det \vb{a} = 0$), then Eq.~\eqref{eq:moebius_trafo} would be just a constant,
\begin{equation} \label{eq:ltfZeroDet}
    \mathcal{L}_{\vb{a}}^{(\det \vb{a} = 0)}(z) = 
    \begin{cases}
    \frac{a_{11}}{a_{10}} & \text{if $a_{10} \neq 0$}\,,\\
    \frac{a_{01}}{a_{00}} & \text{if $a_{10} = 0$ and $a_{00} \neq 0$}\,,\\
    \text{undefined} & \text{if $a_{00} = a_{10} = 0$}\,,
    \end{cases} 
\end{equation}
and thus not strictly considered a M{\"o}bius transformation. $\mathcal{L}_{\vb{a}}(z)$ has the properties
\begin{subequations}\label{eq:ltfProps}
\begin{align}
    \mathcal{L}_{\vb{a}}\left(z \right) &= \mathcal{L}_{\lambda\vb{a}}\left(z \right)\,, \; \text{with} \; \lambda \neq 0\,, \\ 
    \mathcal{L}_{\vb{a}}\left(z \to \infty \right) &= \frac{a_{11}}{a_{10}}\,, \; \text{if the limit exists,}\\
    \mathcal{L}_{\vb{a}}^{-1}\left(z \right) &= \mathcal{L}_{\vb{a}^{-1}}\left(z \right)\,, \; \text{and} \\ 
    \mathcal{L}_{\vb{b}}\left(\mathcal{L}_{\vb{a}}(z) \right) &= \mathcal{L}_{\vb{a}\vb{b}}\left(z \right)\,. 
\end{align}
\end{subequations}
Further, it can be efficiently implemented using (mostly) linear algebra operations; \eg,
\begin{equation}
    \mathcal{L}_{\vb{a}\vb{b}}(z) = \mathcal{F} \left[ \vb{b}^\textsf{T}\vb{a}^\textsf{T}
    \begin{pmatrix}
    \lambda\\\lambda z
    \end{pmatrix}
    \right]
    \qc \text{with} \quad 
        \mathcal{F} \left[ \begin{pmatrix} q\\p \end{pmatrix} \right]
    = \frac{p}{q}\,.
\end{equation}
Note that the vector representation of a fraction is only determined up to an arbitrary factor $\lambda \neq 0$. In this work, we use the M{\"o}bius transformation to relate different asymptotic limit parametrizations with one another.
\section{Solving the radial Schr{\"o}dinger equation}
\label{sec:app:training}

We write the radial Schr{\"o}dinger equation for a given angular momentum $\ell$ and center-of-mass energy $E$ as a system of coupled first-order differential equations,
\begin{equation}
\begin{pmatrix}
    y'_0(r)\\
    y'_1(r)
\end{pmatrix} = 
\begin{pmatrix}
    \phi'(r) \\
    \frac{\ell(\ell+1)}{r^2} + 2\mu \left(V(r; \vb*{\theta}_i)  - E \right) \phi(r)
\end{pmatrix}\,,
\end{equation}
and numerically solve it for each of the $N_b$ partial-wave decomposed potentials $V(r; \vb*{\theta}_1), \ldots$, and $V(r;\vb*{\theta}_{N_b})$ using the explicit Runge-Kutta method in Scipy's \textsc{integrate.solve\_ivp()}. 
The relative and absolute tolerance are each set to $10^{-9}$ or less.
As initial values for the solver we set $\phi(\varepsilon) = 0$ and (by choice) $\phi'(\varepsilon) = 1$, where the value of the derivative will be rescaled later on by imposing an asymptotic boundary condition, and $\varepsilon > 0$ is a numerical value close to zero. We solve the radial Schr{\"o}dinger equation up to the matching radius $r_m \ll \infty$ located outside the range of the potential. At $r_m$, we smoothly match the numerical solution to the free-space solution parametrized by $\phi_{\ell,E}^{\text{(free)}}(r) =  \bar{\phi}_{\ell,E}^{\text{(0,free)}}(r) + L_{\ell,E} \, \bar{\phi}_{\ell,E}^{\text{(1,free)}}(r)$, with 
\begin{equation}\label{eq:freeSolMatch}
\begin{pmatrix}
    \bar{\phi}_{\ell,E}^{\text{(0,free)}}(r)\\
    \bar{\phi}_{\ell,E}^{\text{(1,free)}}(r)
\end{pmatrix} =
pr \, \mathcal{N}^{-1} 
\begin{pmatrix}
    u_{00} & u_{01}\\
    u_{10} & u_{11}\\
\end{pmatrix}
\begin{pmatrix}
     j_\ell(pr) \\
    -\eta_\ell(pr)
\end{pmatrix} \,.
\end{equation}
Here, $j_\ell(pr)$ and $\eta_\ell(pr)$ denote the spherical Bessel function and Neumann function, respectively. Notice that the asymptotic limit of the free-space solution~\eqref{eq:freeSolMatch} is defined in Eq.~\eqref{eq:matrixparam}. The (arbitrary) constant $\mathcal{N}^{-1} = p$ is chosen following Ref.~\cite{Furnstahl:2020abp}. Given a parametrization $\vb{u}$, we determine the value of the $L$-matrix in terms of the inverse logarithmic derivative with respect to $r$, $R(r_m) = \phi(r_m)/\phi'(r_m)$, as follows
\begin{equation}
    L_{\ell,E} = -\frac{\phi_{\ell,E}^{\text{(0,free)}}(r_m) - R(r_m)\phi_{\ell,E}^{\prime \text{(0,free)}}(r_m)}{\phi_{\ell,E}^{\text{(1,free)}}(r_m) - R(r_m)\phi_{\ell,E}^{\prime\text{(1,free)}}(r_m)}\,,
\end{equation}
and then rescale $\phi(r)$ by the factor $\phi_{\ell,E}^\text{(free)}(r_m)/\phi(r_m)$. More details can be found, \eg, in Ref.~\cite{thompson_nunes_2009}.

The numerical solution matched to any asymptotic boundary condition of the form~\eqref{eq:asymLimitParam} is an equally valid solution of the radial Schr{\"o}dinger equation for $r \geqslant \varepsilon $. In practice, we choose a particular boundary condition (\eg, with $L=S$) for the matching. To efficiently transform wave functions normalized by this asymptotic limit parametrization $(L,\vb{u})$ to another $(L',\vb{u}')$, we use the analytic expressions derived in the following. Notice that primes [\eg, as in $\phi'_{0}(r)$] no longer indicate derivatives.

We consider the identity in the asymptotic limit
\begin{equation} \label{eq:matching_L_Lp}
    \phi_{0}(r) + L\, \phi_{1}(r) = C' \left[ \phi'_{0}(r) + L'\, \phi'_{1}(r) \right]\,,
\end{equation}
which implies that $\phi'(r) = C'^{-1}(L) \phi(r)$, and solve for the scalars $C'$ and $L'$ as a function of $L$. For brevity, we omit subscripts that indicate $(E, \ell, \vb*{\theta}_i)$. Equating the coefficients of the sine and cosine functions in Eq.~\eqref{eq:matching_L_Lp} leads to the desired transforms\footnote{Equation~\eqref{eq:LprimeTrafo} can also be obtained by noting that the $K$-matrix~\eqref{eq:trafo_K_L} is independent of whether $(L, \vb{u})$ or $(L', \vb{u}')$ is used to parametrize the asymptotic limit of the radial wave function.}
%
%
%
\begin{align}
    L'(L) &\equiv \mathcal{L}_{\vb{u}'}^{-1}\left(K(L) \right) =  \frac{-u'_{01}+u'_{00} K(L)}{u'_{11} -u'_{10}K(L)}\,, \label{eq:LprimeTrafo} \\ 
    C'(L) &\equiv \mathcal{L}_{\vb{cd}}\left( K(L) \right) = \frac{\det \vb{u}}{\det \vb{u}'} \frac{u'_{11} - u'_{10} K(L)}{u_{11} - u_{10} K(L)}\,,  \label{eq:CprimeTrafo} 
\end{align}
with $K(L)$ as defined in Eq.~\eqref{eq:trafo_K_L},
\begin{equation} \label{eq:def_mat_c_d}
    \vb{c} =
    \begin{pmatrix}
        u_{11} & u'_{11} \\
        -u_{10} & -u'_{10}
    \end{pmatrix}\qc \text{and} \quad
    \vb{d} = 
    \mqty(\dmat[0]{\det \vb{u}', \det \vb{u}})\,. 
\end{equation}
These expressions can be rewritten as $L'(L) = \mathcal{L}_{\vb{u}\vb{u}'^{-1}}(L)$ and $C'(L) = \mathcal{L}_{\vb{ucd}}(L)$, with the generating matrix
\begin{equation} \label{eq:ucd}
    \vb{ucd} = \det \vb{u} \begin{pmatrix}
       \det \vb{u}' & (\vb{u}\vb{u}'^{-1})_{00} \\
       0 & \det \vb{c}  
    \end{pmatrix} \,,
\end{equation}
using the properties~\eqref{eq:ltfProps} of the M{\"o}bius transformation. The generating matrix~\eqref{eq:ucd} is singular if, and only if, $\det \vb{c} \equiv (\vb{u}\vb{u}'^{-1})_{10} = 0$, as expected. In that case, $C'(L) \equiv (\vb{ucd})_{01}/(\vb{ucd})_{00}$ is a constant (\ie, independent of $L$) because of the property~\eqref{eq:ltfZeroDet} and $(\vb{ucd})_{10} = 0$. As discussed in Appendix~\ref{sec:app:redundancies}, this case has important implications for the results obtained from the two KVPs associated with $(L,\vb{u})$ and $(L',\vb{u}')$, respectively. Explicit expressions relating the matrices $(K,S,T)$, as special cases of Eq.~\eqref{eq:LprimeTrafo}, can be found in Table~3.1 of Ref.~\cite{thompson_nunes_2009}.

\section{Efficient evaluation of kernel matrices}
\label{sec:app:kernel}

Constructing the kernel matrix $\Delta \Utilde_{ij}^{(\vb{u})} \propto 2A_{ij} - B_{ij}$ as defined in Eqs.~\eqref{eq:deltaUTilde} for emulating scattering observables with local potentials in coordinate space involves the evaluation of overlap integrals of the functional form
\begin{equation} \label{eq:deltaUdef}
	I_{ij}[V(r;\vb*{\theta})] = \int_0^\infty \dd{r}  \phi_i(r) V(r;\vb*{\theta}) \phi_j(r) \,.
\end{equation}
These integrals can be evaluated to a high accuracy using Gauss--Legendre quadrature rules distributed across multiple intervals. The matrix $B_{ij} = I_{ij}[V(r;\vb*{\theta}_i)+V(r;\vb*{\theta}_j)]$ only depends on the interactions used for training and thus needs to be evaluated only once (for a given $\vb{u}$), whereas $A_{ij} = I_{ij}[V(r;\vb*{\theta})]$ has to be evaluated each time the emulator is invoked after the training phase.

Our emulator applies a set of KVPs with different boundary conditions. Instead of constructing the kernel matrix for each KVP individually, we make use of the analytic transform for the wave functions derived in Appendix~\ref{sec:app:training} to relate two kernel matrices associated with $(L, \vb{u})$ and $(L', \vb{u}')$, respectively. This amounts to the element-wise (\ie, Hadamard) matrix product:
\begin{equation} \label{eq:deltaUTildeTrafo}
    \Delta\Utilde_{ij}^{(\vb{u}')} = C^{\prime-1}(L_i) C^{\prime-1}(L_j) \frac{\det \vb{u}}{\det \vb{u}'} \Delta\Utilde_{ij}^{(\vb{u})} \,,
\end{equation}
where the subscripts index the basis wave functions used for training. While the first two factors on the right-hand side of Eq.~\eqref{eq:deltaUTildeTrafo} transform the wave functions in the integrals of $A_{ij}$ and $B_{ij}$, as discussed in Appendix~\ref{sec:app:training}, the third factor from the left corrects for the different determinants in Eq.~\eqref{eq:deltaUTilde}. Note that a general expression for the inverse of a Hadamard product does not exist. In conclusion, by using the analytic transform~\eqref{eq:deltaUTildeTrafo} combined with Eq.~\eqref{eq:LprimeTrafo} we need to explicitly evaluate the kernel matrix only once each time the emulator is invoked, which allows us to efficiently evaluate an array of different KVPs.

\section{Relationships between Kohn Variational Principles and Kohn anomalies}
\label{sec:app:redundancies}

In this Appendix we inspect the relationship between two arbitrary KVPs associated with $(L, \vb{u})$ and $(L', \vb{u}')$, respectively, and show that their stationary solutions are identical (up to numerical noise) if the cross matrix $\vb{c}$ defined in Eq.~\eqref{eq:def_mat_c_d}, and thus the generating matrix~\eqref{eq:ucd}, is singular (\ie, $\det \vb{c} = \det \vb{ucd} = 0$). For instance, this applies to $(L,L')=(T,S)$ and $(K^{-1},T^{-1})$, as well as combinations drawn from the generalized $T$-matrix KVP,
\begin{align}
    \vb{u}_\tau &= 
    \begin{pmatrix}
    \cos \tau & \sin \tau\\
    -\sin \tau + i \cos \tau & \cos \tau + i \sin \tau
    \end{pmatrix}\,, \label{eq:genTKvp}
    \intertext{and generalized $S$-matrix KVP,}
    \vb{u}_\tau &= 
    \begin{pmatrix}
    - \sin \tau -i\cos \tau & \cos \tau -i \sin \tau\\
    \sin \tau -i\cos \tau &  -\cos \tau -i\sin \tau
    \end{pmatrix}\,, \label{eq:genSKvp}
\end{align}
which reduce to the matrices given in Eq.~\eqref{eq:uMatrices} for the $T$-matrix and $S$-matrix, respectively, when $\tau = 0$. (See also Refs.~\cite{Cooper_2009,Cooper_2010} and the overview of the KVPs in Ref.~\cite{woods2015variational}.)

In Appendix~\ref{sec:app:kernel} we have shown that $C'(L) \equiv C$ is a constant if $\det \vb{c} = 0$. Hence, in that case, the kernel matrices for the two KVPs are directly (not just element-wisely) proportional to one another and Eq.~\eqref{eq:deltaUTildeTrafo} reads
\begin{equation} 
    \Delta\Utilde^{(\vb{u}')} = C \, \Delta\Utilde^{(\vb{u})} \qc \text{with} \quad C = C'^{-2}\frac{\det \vb{u}}{\det \vb{u}'}\,.
\end{equation}
Since Eq.~\eqref{eq:CprimeTrafo}, on the other hand, reduces to
\begin{equation} \label{eq:L_trafo_sing}
    L'(L) = \frac{(\vb{u}\vb{u}'^{-1})_{01}}{(\vb{u}\vb{u}'^{-1})_{00}} + L \, C 
\end{equation}
due to the property~\eqref{eq:ltfZeroDet} of the M{\"o}bius transformation and $(\vb{u}\vb{u}'^{-1})_{10} \equiv \det \vb{c}$, we find that the sets of coefficients $c_i$ for the two KVPs [determined by Eq.~\eqref{eq:kvpSolci}] are equal: the first term from the left in Eq.~\eqref{eq:L_trafo_sing} cancels with the corresponding term from the Lagrange multiplier, while the factor $C$ in the second term cancels with the prefactor $C^{-1}$ of the kernel matrix's inverse. Furthermore, Eqs.~\eqref{eq:deltaUTildeTrafo} and~\eqref{eq:L_trafo_sing} together with $\sum_i c_i = 1$ imply that the stationary approximations for the $L'$-matrix obtained from two KVPs are identical, \ie, 
\begin{equation} \label{eq:id_LLp_kvp}
[L']_\text{KVP} = L'\left([L]_\text{KVP}\right)  \qq{(for $\det \vb{c} = 0$),} 
\end{equation}
where $[L]_\text{KVP}$ is given by Eq.~\eqref{eq:Lapprox}.
For KVPs with $\det \vb{c} \neq 0$, the two approximations are not equal in general, unless the trial wave function is an exact scattering solution, in which case Eq.~\eqref{eq:id_LLp_kvp} is generally fulfilled. 

For real potentials, we moreover find that the coefficients $c_i$ and values of the stationary solutions for two KVPs with $(L,L')=(S,S^{-1})$ are complex conjugated; \ie, $[S]_\text{KVP} = [S^{-1}]^*_\text{KVP}$. This identity follows from $\Delta\Utilde^{(\vb{u}')} = \Delta\Utilde^{*(\vb{u})}$ and $S^{-1} = S^*$ (unitarity), and is generally fulfilled by exact scattering solutions.

Kohn (or Schwartz) anomalies occur at energies $E>0$ where $\det \Delta\Utilde^{(\vb{u})} = 0$ or $\sum_{ij} (\Delta\Utilde^{(\vb{u})})^{-1}_{ij} = 0$.\footnote{Note that the analytic derivation of the stationary approximation~\eqref{eq:kvpSolAna} in Ref.~\cite{Furnstahl:2020abp} assumes that $\sum_{ij} (\Delta\Utilde^{(\vb{u})})^{-1}_{ij}$ in the Lagrange multiplier~\eqref{eq:kvpSolLambda} does not vanish.} In either case, no (unique) stationary approximation for the $L$-matrix can be obtained from the functional~\eqref{eq:betatrial}. Generally, Kohn anomalies reduce the KVP's accuracy over a finite range of the phase space (\eg, the energy) since the stationary solution becomes unstable as $\det \Delta\Utilde^{(\vb{u})}$ or $\sum_{ij} (\Delta\Utilde^{(\vb{u})})^{-1}_{ij}$ becomes vanishingly small (in absolute value). This can be seen in Fig.~\ref{fig:phaseshifts}.

Our findings imply that the KVPs for $(L,L')$ with singular cross matrices $\vb{c}$ [and, for real potentials, also $(L,L')=(S,S^{-1})$] are equally subject to these Kohn anomalies and that deviations from Eq.~\eqref{eq:id_LLp_kvp} are due to numerical noise, \eg, from inverting ill-conditioned kernel matrices.  
Apart from these cases, however, we find for real potentials that complex KVPs (\eg, the $S$-matrix KVP) typically are less prone to Kohn anomalies than real KVPs (\eg, the $K$-matrix KVP) because their kernel matrices are complex,  
which means that both the real and imaginary part of $\det \Delta\Utilde^{(\vb{u})}$ or $\sum_{ij} (\Delta\Utilde^{(\vb{u})})^{-1}_{ij} $ need to approach zero simultaneously. (See also the discussion in Ref.~\cite{Lucchese:1989zz}%
.) For optical potentials, on the other hand, the kernel matrices are complex whether a real or a complex KVP is used.


\section{Diagnostic tools for Kohn anomalies}
\label{sec:app:diagnostic_tools}

As illustrated in Fig.~\ref{fig:phaseshifts}, Kohn anomalies can readily be spotted when the emulated scattering observable of interest is plotted as a function of a continuous variable such as the energy and scattering angle---or likewise when the exact scattering solution is known. Since this is not the case in practice and Monte Carlo sampling can be performed at a fixed energy or scattering angle, a different (\ie, automated) method to detect and remove Kohn anomalies is required for such applications. 

Our method works as follows. For a given $(\ell, E, N_b)$, we simultaneously evaluate the complementary KVPs for $L=(K, T, T^{-1})$ and $L_\tau=(L_{30^\circ}, L_{60^\circ}, L_{90^\circ})$ associated with asymptotic boundary conditions drawn from
\begin{equation} \label{eq:hadinspiredu}
    \vb{u}_\tau = \begin{pmatrix}
    1 & e^{i\tau} \\
    e^{i\tau} & i
    \end{pmatrix}\,, \qq{with} \quad 
    -\frac{\pi}{2} < \tau < \frac{3\pi}{2}\,.
\end{equation} 
The construction and usage of the parameter matrix~\eqref{eq:hadinspiredu} is motivated by the generalized \mbox{$T$-} and $S$-matrix KVP (see Appendix~\ref{sec:app:redundancies} for details), which 
are redundant for EC trial wave functions.
We then compute the relative residuals of, \eg, the estimated $S$-matrices, defined as
\begin{equation} \label{eq:kvp_weight}
    \delta(L_1, L_2) = \max \left\{ \left|  \frac{S(L_1)}{S(L_2)}  - 1 \right|, 
    \left|  \frac{S(L_2)}{S(L_1)}  - 1 \right| \right\}\,,
\end{equation}
for all considered KVPs without repetitions to avoid trivial cases where $L_1=L_2$.\footnote{See also Refs.~\cite{Lucchese:1989zz,Viviani:2001sy,Reeth1996PhD} for similar approaches. If $\delta(L_1, L_2)$ is not well-defined, \eg, $S(L_1)=0$, one could consider the absolute instead of the relative residual in Eq.~\eqref{eq:kvp_weight}.} For instance, given the KVP estimates for $L=(K, T, T^{-1})$ we would determine $\delta(K, T)$, $\delta(K, T^{-1})$, and $\delta(T, T^{-1})$. The expressions for the transformations $S(L)$ are discussed in Appendix~\ref{sec:app:training}. 

Let $\mathcal{P}$ be the set of pairs $(L_1,L_2)$ that fulfill the relative consistency check $\delta(L_1, L_2) < \varepsilon_\text{rel}$, with $\varepsilon_\text{rel} = 10^{-1}$. Since such a consistency check alone does not allow one to disentangle whether $L_1$, $L_2$, or both are anomalous, we estimate the $S$-matrix by the weighted sum of averages
\begin{subequations} \label{eq:mix_av_sum}
\begin{align} 
    [S]_\text{KVP}^\text{(mixed)} &= \sum_{(L_1,L_2) \in \mathcal{P}} 
    \omega(L_1, L_2)
    \frac{S(L_1) + S(L_2)}{2} \,,\\
    \omega(L_1, L_2) &= \frac{\delta(L_1, L_2)^{-1}}{\sum_{(L'_1,L'_2) \in \mathcal{P}} \delta(L'_1, L'_2)^{-1}}\,,
\end{align}
\end{subequations}
if at least one consistency check passes (see also Ref.~\cite{ARMOUR1991165}). A small regularization parameter is added to $\delta(L_1, L_2)$ to prevent a potential division by zero.
If all checks fail, we partition the training set with $N_b$ wave functions in batches of size $N_p \leqslant N_b$,\footnote{The size of the last batch will be slightly larger than $N_p$ if the integer division $N_b/N_p$ has a nonzero remainder.} remove one batch at a time from the training set (\ie, then of size $N_b-N_p$), and repeat the process iteratively. 
This usually shifts the Kohn anomalies in each iteration.  
The iterative process is computationally efficient because $\Delta\Utilde^{(\vb{u})}$ only needs to be sliced, not recomputed. Different ways of partitioning the training set can be straightforwardly implemented.

For instance, suppose $N_b = 6$ and $N_p = 3$, we would first remove the basis wave functions with indices $i = (1,2,3)$ and run the consistency checks; if all checks fail again, we add the basis wave functions with $i = (1,2,3)$ back to the training sets and remove the ones with $i = (4,5,6)$ next. If all checks fail once more, our algorithm signals that Kohn anomalies could not be mitigated for the given training set and terminates. This can happen in practice, \eg, if all KVPs considered are anomalous at overlapping regions in the phase space. 

\section{Additional results} \label{sec:app:add_results}

\begin{figure*}[tb]
\begin{centering}
\includegraphics[width=0.7\linewidth]{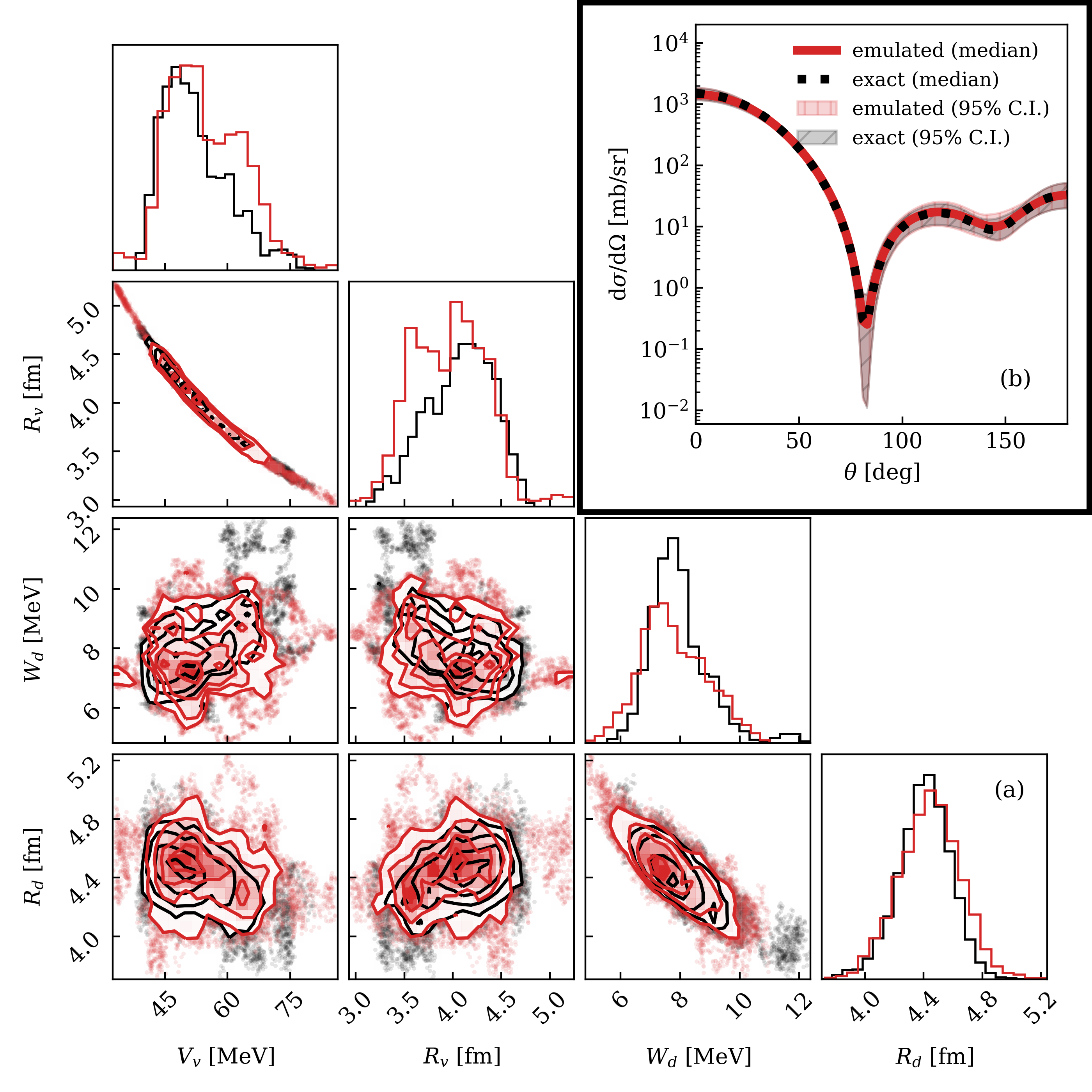}
\end{centering}
\caption{Comparison of the MCMC results obtained for the mixed approach using $N_b=8$ (red) and the exact solution (black) for \isotope[40]{Ca}$(n,n)$ scattering at $5 \MeV$: (a)~posterior distributions for the parameters $V_v, R_v, W_d$, and $R_d$ along the diagonal, and correlations between each pair of parameters in the off-diagonal; (b)~the median value of the differential cross section as a function of scattering angle (lines) and the corresponding 95\% confidence interval (C.I., shaded area). These MCMC calculations correspond to 20,000 accepted parameters sets.}
\label{fig:mcmc40Ca}
\end{figure*}

\begin{table}[tb]
    \caption{Mean values ($\overline{p}$) and standard deviations ($\Delta p$) for the four parameters varied in the MCMC optimization of $^{40}$Ca$(n,n)$ scattering at $5\MeV$ as obtained from the EC-driven emulator (mixed approach) and the exact scattering solution. Note that the values for the radii are given for $R_i=r_iA^{1/3}$.  
    }
    \label{tab:40CanParms}
    \centering
    \begin{ruledtabular}
    \begin{tabular}{ldddd}
    \multicolumn{1}{c}{} & \multicolumn{2}{c}{emulated} & \multicolumn{2}{c}{exact} \\
    Parameter & \multicolumn{1}{c}{$\overline{p}$} & \multicolumn{1}{c}{$\Delta p$} & \multicolumn{1}{c}{$\overline{p}$} & \multicolumn{1}{c}{$\Delta p$} \\ \hline
$V_v$ [MeV] & 54.82 & 8.95 & 52.39 & 7.94 \\
$R_v$ [fm] & 3.97 & 0.38 & 4.07 & 0.34 \\
$W_v$ [MeV] & 7.79 & 1.11 & 7.97 & 1.00 \\
$R_w$ [fm] & 4.44 & 0.21 & 4.40 & 0.18  \\
    \end{tabular}
    \end{ruledtabular}
\end{table}

We provide here additional results for $^{40}$Ca$(n,n)$ scattering at $E=5 \MeV$.
Figure~\ref{fig:mcmc40Ca} shows the posterior distributions for the four varied parameters, similar to Fig.~\ref{fig:mcmc} but at the lower energy.
The differential cross sections and 95\% confidence intervals for the exact scattering solution and emulator are in good agreement, as are the means and standard deviations (see Table~\ref{tab:40CanParms}), and the correlations of the parameters.

\bibliography{bib, bayesian_refs}

\end{document}